\documentclass[12pt]{article}
\def\beq{\begin{equation}}
\def\eeq{\end{equation}}
\def\bea{\begin{eqnarray}}
\def\eea{\end{eqnarray}}
\def\nn{\nonumber}
\def\sss{\scriptscriptstyle}
\def\bd{B_d^0}
\def\bdbar{{\overline{B_d^0}}}

\def\barp{{\raise.35ex\hbox
{${\sss (}$}}---{\raise.35ex\hbox{${\sss )}$}}}
\def\bdbarp{\hbox{$B_d$\kern-1.4em\raise1.4ex\hbox{\barp}}}
\def\bsbarp{\hbox{$B_s$\kern-1.4em\raise1.4ex\hbox{\barp}}}
\def\ks{K_{\sss S}}

\def\roughly#1{\mathrel{\raise.3ex\hbox
{$#1$\kern-.75em\lower1ex\hbox{$\sim$}}}}

\def\tnp{\theta_{\sss NP}}
\def\alphaeff{{\alpha_{\sss eff}}}

\def\npb#1#2#3{{ Nucl.\ Phys.} {\bf B#1}, #3 (#2)}
\def\plb#1#2#3{{ Phys.\ Lett.} {\bf #1B}, #3 (#2)}
\def\prd#1#2#3{{ Phys.\ Rev.} {\bf D#1}, #3 (#2)}
\def\prl#1#2#3{{ Phys.\ Rev.\ Lett.} {\bf #1}, #3 (#2)}

\newread\epsffilein 
\newif\ifepsffileok 
\newif\ifepsfbbfound 
\newif\ifepsfverbose 
\newdimen\epsfxsize 
\newdimen\epsfysize 
\newdimen\epsftsize 
\newdimen\epsfrsize 
\newdimen\epsftmp 
\newdimen\pspoints 
\def\epsfbox#1{\global\def\epsfllx{72}\global\def\epsflly{72}%
 \global\def\epsfurx{540}\global\def\epsfury{720}%
 \def\lbracket{[}\def\testit{#1}\ifx\testit\lbracket
 \let\next=\epsfgetlitbb\else\let\next=\epsfnormal\fi\next{#1}}%
\def\epsfgetlitbb#1#2 #3 #4 #5]#6{\epsfgrab #2 #3 #4 #5 .\\%
 \epsfsetgraph{#6}}%
\def\epsfnormal#1{\epsfgetbb{#1}\epsfsetgraph{#1}}%
\def\epsfgetbb#1{%
\openin\epsffilein=#1
\ifeof\epsffilein\errmessage{I couldn't open #1, will ignore it}\else
 {\epsffileoktrue \chardef\other=12
 \def\do##1{\catcode`##1=\other}\dospecials \catcode`\ =10
 \loop
 \read\epsffilein to \epsffileline
 \ifeof\epsffilein\epsffileokfalse\else
 \expandafter\epsfaux\epsffileline:. \\%
 \fi
 \ifepsffileok\repeat
 \ifepsfbbfound\else
 \ifepsfverbose\message{No bounding box comment in #1; using defaults}\fi\fi
 }\closein\epsffilein\fi}%
\def\epsfclipstring{}
\def\epsfsetgraph#1{%
 \epsfrsize=\epsfury\pspoints
 \advance\epsfrsize by-\epsflly\pspoints
 \epsftsize=\epsfurx\pspoints
 \advance\epsftsize by-\epsfllx\pspoints
 \epsfxsize\epsfsize\epsftsize\epsfrsize
 \ifnum\epsfxsize=0 \ifnum\epsfysize=0
 \epsfxsize=\epsftsize \epsfysize=\epsfrsize
 \epsfrsize=0pt
 \else\epsftmp=\epsftsize \divide\epsftmp\epsfrsize
 \epsfxsize=\epsfysize \multiply\epsfxsize\epsftmp
 \multiply\epsftmp\epsfrsize \advance\epsftsize-\epsftmp
 \epsftmp=\epsfysize
 \loop \advance\epsftsize\epsftsize \divide\epsftmp 2
 \ifnum\epsftmp>0
 \ifnum\epsftsize<\epsfrsize\else
 \advance\epsftsize-\epsfrsize \advance\epsfxsize\epsftmp \fi
 \repeat
 \epsfrsize=0pt
 \fi
 \else \ifnum\epsfysize=0
 \epsftmp=\epsfrsize \divide\epsftmp\epsftsize
 \epsfysize=\epsfxsize \multiply\epsfysize\epsftmp
 \multiply\epsftmp\epsftsize \advance\epsfrsize-\epsftmp
 \epsftmp=\epsfxsize
 \loop \advance\epsfrsize\epsfrsize \divide\epsftmp 2
 \ifnum\epsftmp>0
 \ifnum\epsfrsize<\epsftsize\else
 \advance\epsfrsize-\epsftsize \advance\epsfysize\epsftmp \fi
 \repeat
 \epsfrsize=0pt
 \else
 \epsfrsize=\epsfysize
 \fi
 \fi
 \ifepsfverbose\message{#1: width=\the\epsfxsize, height=\the\epsfysize}\fi
 \epsftmp=10\epsfxsize \divide\epsftmp\pspoints
 \vbox to\epsfysize{\vfil\hbox to\epsfxsize{%
 \ifnum\epsfrsize=0\relax
 \includegraphics{#1}%
 \else
 \epsfrsize=10\epsfysize \divide\epsfrsize\pspoints
 \includegraphics{#1}%
 \fi
 \hfil}}%
\global\epsfxsize=0pt\global\epsfysize=0pt}%
 {\catcode`\%=12 \global\let\epsfpercent=
\long\def\epsfaux#1#2:#3\\{\ifx#1\epsfpercent
 \def\testit{#2}\ifx\testit\epsfbblit
 \epsfgrab #3 . . . \\%
 \epsffileokfalse
 \global\epsfbbfoundtrue
 \fi\else\ifx#1\par\else\epsffileokfalse\fi\fi}%
\def\epsfempty{}%
\def\epsfgrab #1 #2 #3 #4 #5\\{%
\global\def\epsfllx{#1}\ifx\epsfllx\epsfempty
 \epsfgrab #2 #3 #4 #5 .\\\else
 \global\def\epsflly{#2}%
 \global\def\epsfurx{#3}\global\def\epsfury{#4}\fi}%
\def\epsfsize#1#2{\epsfxsize}
\pagestyle{plain}
\begin{document}

\begin{flushright}  
UdeM-GPP-TH-03-116\\
McGill/03-27
\end{flushright}

\begin{center}
\bigskip
{\Large \bf CP Violation in $B\to\rho\pi$: New Physics Signals} \\
\bigskip
{\large V\'eronique Pag\'e
$^{a,}$\footnote{veronique.page@umontreal.ca} and David London
$^{a,b,}$\footnote{london@lps.umontreal.ca}} \\
\end{center}

\begin{flushleft}
~~~~~~~~~~~$a$: {\it Laboratoire Ren\'e J.-A. L\'evesque, 
Universit\'e de Montr\'eal,}\\
~~~~~~~~~~~~~~~{\it C.P. 6128, succ. centre-ville, Montr\'eal, QC,
Canada H3C 3J7\\
~~~~~~~~~~~$b$: {\it Physics Department, McGill University,}\\
~~~~~~~~~~~~~~~{\it 3600 University St., Montr\'eal QC, Canada H3A 2T8}}
\end{flushleft}

\begin{center} 
\bigskip (\today)
\vskip0.5cm
{\Large Abstract\\}
\vskip3truemm
\parbox[t]{\textwidth} {A Dalitz-plot analysis of $\bd(t) \to \rho\pi
\to \pi^+\pi^-\pi^0$ decays allows one to obtain the CP-violating
phase $\alpha$. In addition, one can extract the various tree ($T$)
and penguin ($P$) amplitudes contributing to these decays. By
comparing the measured value of $|P/T|$ with the theoretical
prediction, one can detect the presence of physics beyond the standard
model.}
\end{center}

\thispagestyle{empty}
\newpage
\setcounter{page}{1}
\baselineskip=14pt

A great many methods have been proposed for obtaining information
about the CP phases $\alpha$, $\beta$ and $\gamma$ of the unitarity
triangle \cite{pdg}. Almost all of these involve CP-violating
asymmetries in hadronic $B$ decays \cite{CPreview}. The aim is to test
the standard model (SM) explanation of CP violation, and hopefully
find evidence for physics beyond the SM.

The cleanest methods (i.e.\ those in which the theoretical hadronic
uncertainties are very small) involve $B$ decays which are dominated
by a single amplitude, such as $\bd(t) \to J/\psi \ks$. However, many
decays receive contributions from both tree and penguin diagrams with
different weak phases \cite{penguins}. A-priori, one would think that
one cannot obtain clean phase information from such decays.
Fortunately, techniques have been developed for removing the unwanted
``penguin pollution.'' For example, an isospin analysis of
$B\to\pi\pi$ decays allows one to remove this contamination and obtain
$\sin 2\alpha$ cleanly \cite{isospin}, albeit with discrete
ambiguities.

In fact, this isospin analysis gives us even more information. In
particular, one can also obtain the magnitudes and relative phases of
the tree ($T$) and penguin ($P$) amplitudes in $\bd\to\pi^+\pi^-$
\cite{Charles}. It is therefore possible to compare the experimental
value of $|P/T|$ with that predicted by theory. If a significant
discrepancy is observed, it would signal new physics \cite{LSSpipi}.

An alternative technique for obtaining $\alpha$ involves $B\to\rho\pi$
decays. By performing a Dalitz-plot analysis of $\bd(t) \to \rho\pi
\to \pi^+\pi^-\pi^0$ decays, one can remove the penguin contributions
from $B\to\rho\pi$ decays and obtain $\alpha$ \cite{Dalitz}. Compared
to $B\to\pi\pi$, the advantage of this method is that it is possible
to extract both $\sin 2\alpha$ and $\cos 2\alpha$, so that one obtains
$2\alpha$ with no discrete ambiguity. Another advantage is that it is
not necessary to measure processes involving two final-state $\pi^0$
mesons. The disadvantage of this method is that one must understand
the continuum background to such decays with considerable accuracy, as
well as the correct description of $\rho\to\pi\pi$ decays, and these
may be difficult.

Here too there is enough information to obtain the magnitudes and
relative phases of the tree and penguin amplitudes. Thus, one can
measure $|P/T|$ in $B\to\rho\pi$. As in $B\to\pi\pi$, a comparison of
this ratio with the theoretical prediction can reveal the presence of
new physics. In this paper we perform such an analysis. As we will
show, the $B\to\rho\pi$ method has two advantages compared to
$B\to\pi\pi$ for searching for physics beyond the SM in this
way. First, the fact that there is no discrete ambiguity in $2\alpha$
improves the prospects for finding new physics. Second, the $|P/T|$
ratio is expected to be smaller than in $B\to\pi\pi$, which makes it
easier to see a new-physics signal, should it be present.

We begin with a brief review of the $B\to\rho\pi$ Dalitz-plot analysis
within the SM \cite{Dalitz}. There are five $B\to\rho\pi$ amplitudes
which satisfy a pentagon isospin relation. All amplitudes receive
contributions from both tree and ${\bar b}\to {\bar d}$ penguin
amplitudes. The tree amplitude is proportional to $V_{ub}^* V_{ud}$,
while the penguin amplitude has contributions from internal $u$, $c$
and $t$ quarks, proportional to $V_{ub}^* V_{ud}$, $V_{cb}^* V_{cd}$
and $V_{tb}^* V_{td}$, respectively. Using the unitarity of the CKM
matrix, $V_{td}^*V_{tb} + V_{cd}^*V_{cb} + V_{ud}^*V_{ub} = 0$, we can
eliminate the $c$-quark contribution. Furthermore, the piece
proportional to $V_{ub}^* V_{ud}$ can be absorbed into the tree
amplitude. Thus, the penguin amplitude includes only the piece
proportional to $V_{td}^*V_{tb}$. It is convenient to rescale the
amplitudes by $e^{i\beta}$, leading to the following expressions for
the amplitudes:
\bea
\label{SMamps}
S_{+0} &\!\equiv\!& e^{i\beta} \sqrt{2} A(B^+ \to \rho^+ \pi^0) =
T^{+0}e^{-i\alpha} + P^{+0} ~, \nn\\
S_{0+} &\!\equiv\!& e^{i\beta} \sqrt{2} A(B^+ \to \rho^0 \pi^+) =
T^{0+}e^{-i\alpha} + P^{0+} ~, \nn\\
S_{+-} &\!\equiv\!& e^{i\beta} A(B^0 \to \rho^+ \pi^-) =
T^{+-}e^{-i\alpha} + P^{+-} ~, \\
S_{-+} &\!\equiv\!& e^{i\beta} A(B^0 \to \rho^- \pi^+) =
T^{-+}e^{-i\alpha} + P^{-+} ~, \nn\\
S_{00} &\!\equiv\!& e^{i\beta} 2 A(B^0 \to \rho^0 \pi^0) = S_{+0} + S_{0+}
- S_{+-} - S_{-+} ~, \nn
\eea
where $P^{0+} = - P^{+0}$. In the above we have explicitly written the
weak phase $\alpha$, while the $T_i$ and the $P_i$ include strong
phases. (Throughout the paper, we use the subscript `$i$' to denote
all of the $\rho\pi$ charge combinations: $i = +0$, $0+$, $+-$, $-+$,
$00$.) The corresponding amplitudes for the CP-conjugate processes,
${\bar S}_i$, are obtained by changing the signs of the weak phases.

The key point is that all of the neutral $\bd\to\rho\pi$ amplitudes
contribute to $\bd\to\pi^+\pi^-\pi^0$. We can therefore write
\beq
A(\bd\to\pi^+\pi^-\pi^0)= f^+ S_{+-} + f^- S_{-+} + f^0 S_{00} / 2 ~,
\label{Bdamp}
\eeq
where the $f^i$ are the kinematic distribution functions for the pions
produced in the decay of the $\rho^i$. The $\bdbar$ mesons can decay
to the same final state:
\beq
A(\bdbar\to\pi^+\pi^-\pi^0)= f^- {\bar S}_{+-} + f^+ {\bar S}_{-+} +
f^0 {\bar S}_{00} / 2 ~.
\label{Bdbaramp}
\eeq
The time-dependent measurement of the Dalitz plot for $\bd(t)\to
\pi^+\pi^-\pi^0$ then allows one to extract the magnitudes and
relative phases of each of the $f^i$, $S_{ij}$ and ${\bar S}_{ij}$ in
Eqs.~(\ref{Bdamp}) and (\ref{Bdbaramp}) \cite{Dalitz}. By taking the
ratio of the relations
\bea
S_{+-} + S_{-+} + S_{00} & = & (T^{+0} + T^{0+}) \, e^{-i\alpha} ~,
\nn\\
{\bar S}_{+-} + {\bar S}_{-+} + {\bar S}_{00} & = & (T^{+0} + T^{0+})
\, e^{i\alpha} ~,
\eea
one obtains $e^{-2 i\alpha}$. We therefore see that, using this
method, the CP phase $2\alpha$ can be extracted with no ambiguity.

It is also possible to obtain the tree and penguin contributions to
the amplitudes in Eq.~(\ref{SMamps}). We define the following
observables:
\bea
B_i & \equiv & {1 \over 2} (|A_i|^2 +|{\bar A_i}|^2) ~, \nn\\
a_i & \equiv & {|S_i|^2 - |{\bar S_i}|^2 \over |S_i|^2 + |{\bar
S_i}|^2} ~, \nn\\
2\alpha_{eff}^i & \equiv & {\rm Arg}({\bar S}_iS_i^*) ~.
\label{observables}
\eea
Here $B_i$, $a_i$ and $2\alpha_{eff}^i$ are, respectively, the
branching ratio, direct CP asymmetry, and measure of indirect CP
violation for each decay. We remark that each of $2\alpha_{eff}^{+0}$
and $2\alpha_{eff}^{0+}$ are automatically zero since they involve
charged $B$ decays. (Note: the indirect CP asymmetry is usually
written with an explicit mixing phase $q/p = e^{-2i\beta}$. This phase
is removed when one rescales the amplitudes by $e^{i\beta}$ as in
Eq.~(\ref{SMamps}).)  We have
\bea
S_i - {\bar S_i} & = & -2i\sin \alpha \, T_i ~, \nn\\
S_i \, e^{i\alpha} - {\bar S_i} \, e^{-i\alpha} & = & -2i\sin \alpha
\, P_i ~.
\eea
It is then straightfoward to obtain $|T_i|^2$ and $|P_i|^2$:
\bea
|T_i|^2 & = & R_i \, {1 - \sqrt{1-a_i^2} \cos 2\alpha_{eff}^i \over 1 -
  \cos 2\alpha} ~, \nn\\
|P_i|^2 & = & R_i \, {1 - \sqrt{1-a_i^2} \cos (2\alpha_{eff}^i -
  2\alpha) \over 1 - \cos 2\alpha} ~.
\eea
where
\beq
R_i \equiv {(|S_i|^2 + |{\bar S}_i|^2) \over 2} ~.
\eeq
Note that $R_i$ is proportional to $B_i$ [Eq.~(\ref{observables})].
The proportionality constant depends on which decay is being
considered, see Eq.~(\ref{SMamps}).

Suppose now that there is physics beyond the SM. If present, it will
affect mainly $\bd$--$\bdbar$ mixing and/or the ${\bar b}\to {\bar d}$
penguin amplitude. In the SM, the weak phase of $\bd$--$\bdbar$ mixing
($\beta$) is equal to that of the $t$-quark contribution to the ${\bar
b}\to {\bar d}$ penguin. This is reflected in the fact that the weak
phase multiplying the term $P_i$ in Eq.~(\ref{SMamps}) is zero. If new
physics is present, these two weak phases may be different. One can
take this possibility into account by including a new-physics phase
$\tnp$ in the $B\to\rho\pi$ amplitudes:
\bea
S_{+0}  & = & T^{+0}e^{-i\alpha} + P^{+0}e^{-i\tnp} ~, \nn\\
S_{0+}  & = & T^{0+}e^{-i\alpha} + P^{0+}e^{-i\tnp} ~, \nn\\
S_{+-}  & = & T^{+-}e^{-i\alpha} + P^{+-}e^{-i\tnp} ~, \nn\\
S_{-+}  & = & T^{-+}e^{-i\alpha} + P^{-+}e^{-i\tnp} ~, \nn\\
S_{00}  & = & S_{+0} + S_{0+} - S_{+-} - S_{-+}  ~.
\label{NPamps}
\eea
The extraction of $\alpha$ is unchanged by the presence of the
new-physics parameter $\tnp$ (though its value may include new
contributions to $\bd$--$\bdbar$ mixing). However, the expressions for
$T_i$ and $P_i$ are modified. We now have
\bea
S_i \, e^{i\tnp} - {\bar S_i} \, e^{-i\tnp} & = &
-2i\sin(\alpha - \tnp) \, T_i ~, \nn\\
S_i \, e^{i\alpha} - {\bar S_i} \, e^{-i\alpha} & = &
-2i\sin(\alpha - \tnp) \, P_i ~,
\eea
so that
\bea
|T_i|^2 & = & R_i \, {1 - \sqrt{1-a_i^2} \cos (2\alpha_{eff}^i - 2\tnp)
  \over 1 - \cos (2\alpha -2\tnp)}~, \nn\\
|P_i|^2 & = & R_i \, {1 - \sqrt{1-a_i^2} \cos (2\alpha_{eff}^i -
  2\alpha) \over 1 - \cos (2\alpha -2\tnp)} ~.
\eea
The expressions for the $T_i$ and $P_i$ are therefore altered in the
presence of new physics. Thus, by comparing the measured value of a
particular $|P/T|$ with that predicted by theory (within the SM), we
can detect the presence of a nonzero $\tnp$. (It is also possible for
new physics to affect the magnitudes of the $T_i$ and $P_i$. This
possibility is implicitly included in our method.)

The first step is therefore to evaluate the theoretical value of
$|P/T|$. However, there are many $|P/T|$ ratios that can be
considered. We concentrate only on the (color-allowed) neutral decays
$\bd \to \rho^\pm \pi^\mp$. There are several reasons for this. First,
the Dalitz plots for the charged $B$ decays are much more difficult to
obtain since they require the detection of two $\pi^0$'s. Second, the
branching ratio for the color-suppressed decay $\bd \to \rho^0\pi^0$
is expected to be quite a bit smaller than those of $\bd \to \rho^\pm
\pi^\mp$. Finally, below we will use QCD factorization to estimate the
theoretical size of the $|P/T|$ ratios, and nonfactorizable effects
are expected to be small for color-allowed decays.

The value of $|P/T|$ for $\bd \to \rho^\pm \pi^\mp$ has been
calculated in the literature. In Ref.~\cite{Aleksan}, this was done
using naive factorization and including only the $t$-quark
contribution to the ${\bar b} \to {\bar d}$ penguin. A more recent
computation has been done by Beneke and Neubert \cite{BN} in the
context of QCD factorization \cite{BBNS}. Since QCD factorization is a
state-of-the-art framework, using expansions in $1/m_b$ and
$\alpha_s$, we will follow this approach. (It should be noted,
however, that the $|P/T|$ ranges given in Refs.~\cite{Aleksan} and
\cite{BN} are quite similar.)

We define
\beq
r^{+-} \equiv \left\vert {P^{+-} \over T^{+-}} \right\vert ~,~~
r^{-+} \equiv \left\vert {P^{-+} \over T^{-+}} \right\vert ~.
\label{PTratio}
\eeq
Ref.~\cite{BN} gives
\beq
r^{+-} = 0.10^{+0.06}_{-0.04} ~,~~
r^{-+} = 0.10^{+0.09}_{-0.05} ~.
\label{BNrranges}
\eeq
The errors come principally from three sources: the values of
$|V_{ub}|$ and $m_s$, and the size of weak annihilation effects. Note
that the two ratios are determined by very different dynamics, so that
their near equality is a numerical accident.

It is now necessary to decide on the numerical ranges to use for
$r^{+-}$ and $r^{-+}$ in the analysis. Since the goal is to search for
physics beyond the SM, it is important to be as conservative as
possible. With this in mind, we will take the theoretical ranges for
$r^{+-}$ and $r^{-+}$ within the SM to be
\beq
0.05 < r^{+-} < 0.25 ~,~~ 0.05 < r^{-+} < 0.25 ~.
\label{rrange}
\eeq
The above ranges are larger than those given in Eq.~(\ref{BNrranges}),
particularly on the upper side. We note that QCD factorization cannot
account for the observed $B\to\pi\pi$ and $B\to\rho\pi$ branching
ratios \cite{BN}. Assuming no new physics --- and the analysis of this
paper can be used to test for such effects --- there must be some
contribution which is larger than its QCD factorization value. The
enlarged ranges of Eq.~(\ref{rrange}) take this into account, as well
as potential underestimates of factorizable errors (e.g.\
electroweak-penguin effects) and nonfactorizable effects.  With the
ranges of Eq.~(\ref{rrange}), a significant discrepancy between the
measured value of $|P/T|$ and its SM prediction will clearly be a sign
of new physics. That is, within the SM, we expect
\beq
0.05 < \sqrt{ 1- \sqrt {1-a_i^2} \cos (2\alpha_{eff}^i - 2\alpha)
  \over 1 - \sqrt{1-a_i^2} \cos 2\alpha_{eff}^i } < 0.25 ~,
\eeq
for $i = +-$, $-+$. If it is found that the observables do not respect
this inequality, this points to the presence of physics beyond the SM.

We note in passing that, while the $|P/T|$ range in $B\to\rho\pi$ is
$\sim 10\%$, in $B\to\pi\pi$ it is expected to be $\sim 20$--30\%
\cite{LSSpipi,BN,FM}. We therefore conclude that the penguin pollution
is likely to be more significant in $B\to\pi\pi$ \cite{Aleksan}.
Thus, if new physics is present, it will be easier to detect in
$B\to\rho\pi$.

As noted earlier, the CP phase $2\alpha$ can be extracted from the
$B\to\rho\pi$ method. However, this is not easy experimentally. In our
analysis we therefore consider the possibility that only $\sin
2\alpha$ is measured (in which case one obtains $2\alpha$ with a
twofold ambiguity), as well as the case where $2\alpha$ is known
without ambiguity. In both scenarios, we consider two possible ranges
for $2\alpha$: (i) $120^\circ \le 2\alpha \le 135^\circ$, (ii)
$165^\circ \le 2\alpha \le 180^\circ$, which can be considered to take
into account the experimental errors in the measurements.

\begin{figure*}
\centerline{\epsfxsize 7.0truein \epsfbox {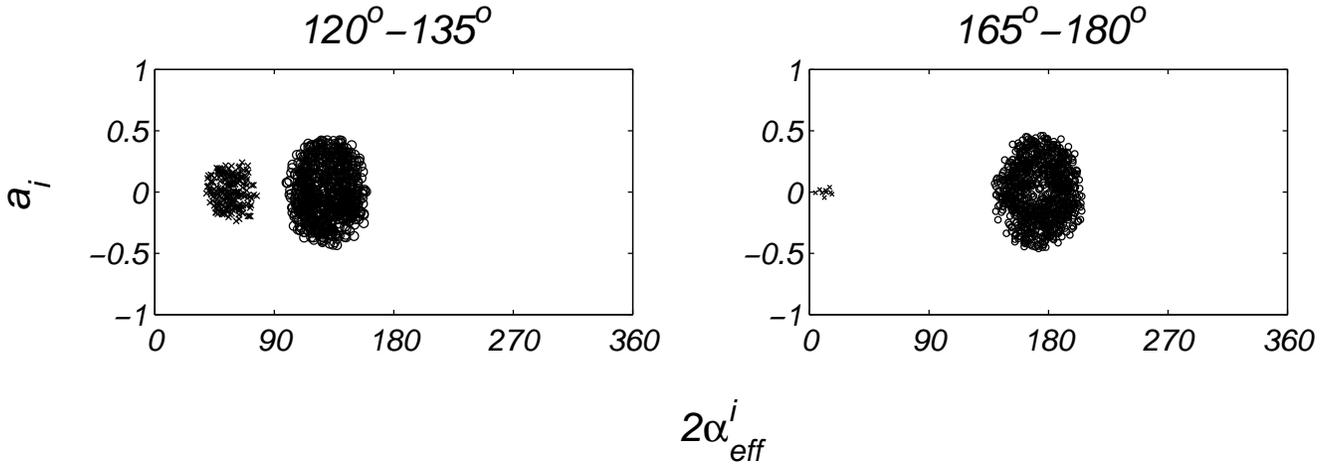}}
\caption{The region in $2\alphaeff^i$--$a_i$ space ($i=+-$, $-+$)
which is consistent with the theoretical prediction for $|P^i/T^i|$
[Eqs.~(\protect\ref{PTratio}), (\protect\ref{rrange})], for two ranges
of $2\alpha$, shown above each figure. It is assumed that only $\sin
2\alpha$ is measured, so that $2\alpha$ is obtained with a twofold
ambiguity. (This is the source of the two regions in both figures.)
If $\cos 2\alpha$ is also measured, allowing one to obtain $2\alpha$
without ambiguity, the left-hand region must be removed in both
figures.}
\label{Fig1}
\end{figure*}

Our results are shown in Fig.~\ref{Fig1}. We consider the two ranges
for $2\alpha$ given above. For each of these ranges, Fig.~\ref{Fig1}
shows the (correlated) allowed values of $2\alpha_{eff}^i$ and $a_i$
($i=+-$, $-+$) that are consistent, within the SM, with the assumed
range for $2\alpha$ and the theoretical range for $|P^i/T^i|$
[Eq.~(\ref{rrange})]. If only $\sin 2\alpha$ has been measured, then,
for a given range of $2\alpha$, both regions in Fig.~\ref{Fig1} are
allowed. If $\cos 2\alpha$ can also be measured, then the left-hand
region can be removed. In either of these scenarios, if the measured
values of the observables do not lie within the SM region, this means
that new physics --- i.e.\ a nonzero $\tnp$ --- is present. As can be
seen from Fig.~\ref{Fig1}, the new-physics region is quite large, so
that we have a good chance of detecting the new physics via this
method, should it be present.

If no signal for new physics is detected, one can place an upper limit
on the value of $\tnp$ via
\beq
0.05 < \sqrt{ 1- \sqrt {1-a_i^2} \cos (2\alpha_{eff}^i -
  2\alpha) \over 1 - \sqrt{1-a_i^2} \cos (2\alpha_{eff}^i - 2\tnp)}
  < 0.25 ~.
\eeq

To summarize, the measurement of the Dalitz plot of $\bd(t) \to
\rho\pi \to \pi^+\pi^-\pi^0$ decays allows one to cleanly extract the
CP-violating phase $2\alpha$, with no discrete ambiguity. One can also
obtain the individual tree ($T$) and penguin ($P$) amplitudes in these
decays. By comparing the measured value of a particular $|P/T|$ ratio
with that predicted by theory, one can detect the presence of physics
beyond the standard model. From both a theoretical and experimental
point of view, the best $|P/T|$ ratios are those for $\bd \to \rho^\pm
\pi^\mp$. The conservative ranges for these ratios are taken to be
$0.05 < \left\vert P^i/T^i \right\vert < 0.25$ ($i=+-$, $-+$). The
region in $(2\alpha_{eff}^i,a_i)$ parameter space ($2\alpha_{eff}^i$
and $a_i$ are, respectively, the measured indirect and direct CP
asymmetries) which corresponds to this range of $\left\vert P^i/T^i
\right\vert$ is relatively small. This therefore provides a good way
of detecting the presence of new physics.

\section*{\bf Acknowledgments}

We thank S. Hikspoors for collaboration in the initial stages of this
project and A. H\"ocker for helpful communications. This work was
financially supported by NSERC of Canada.

\end{document}